\documentclass[runningheads]{llncs}

\usepackage[T1]{fontenc}

\usepackage{amsmath,amssymb,amsfonts,mathrsfs}
\usepackage{multicol}

\usepackage[linkcolor=black,colorlinks=true,citecolor=black,filecolor=black,draft=false]{hyperref}
\usepackage[nameinlink]{cleveref}
\Crefname{figure}{Figure}{Figures}

\usepackage{color}

\usepackage{tikz}
\newsavebox{\fminipagebox}
\NewDocumentEnvironment{fminipage}{m O{\fboxsep}}
{\par\kern#2\noindent\begin{lrbox}{\fminipagebox}
        \begin{minipage}{#1}\ignorespaces}
        {\end{minipage}\end{lrbox}%
    \makebox[#1]{%
        \kern\dimexpr-\fboxsep-\fboxrule\relax
        \fbox{\usebox{\fminipagebox}}%
        \kern\dimexpr-\fboxsep-\fboxrule\relax
    }\par\kern#2
}

\newcounter{protocol}
\crefname{protocol}{Protocol}{Protocols}

\newenvironment{protocol}[2]
{%
    \refstepcounter{protocol}
    \def\protocolendargument{#1}
    \begin{figure}[!h]
        \label{#2}
        \noindent
        \begin{fminipage}{\textwidth}
        }%
        {%
        \end{fminipage}
        {\textbf{Protocol \theprotocol:} \protocolendargument}
    \end{figure}
}

\newcommand{\customarrow}[1]{\begin{tikzpicture}[baseline=(current bounding box.base)]\draw [#1](0,0) -- (1.2,0);\end{tikzpicture}}
\newcommand{\customarrownamed}[3]{\begin{tikzpicture}[baseline=(current bounding box.base)]\draw [#1](0,0) -- (1.2,0);\node at (0.18,-0.14) {\tiny #2};\node at (1,-0.14) {\tiny #3};\end{tikzpicture}}

\newcommand{\customarrownamedtext}[3]{\begin{tikzpicture}[baseline={([yshift=-.8ex]current bounding box.center)}]\draw [#1](0,0) -- (0.6,0);\node at (0.18,-0.14) {\tiny #2};\node at (0.41,-0.14) {\tiny #3};\end{tikzpicture}}

\newcommand{\sendleft}{\customarrow{stealth-}}

\newcommand{\sendleftsecure}{\customarrownamed{stealth-}{S}{ }}

\newcommand{\sendright}{\customarrow{-stealth}}
\newcommand{\sendrightauth}{\customarrownamed{-stealth}{ }{A}}

\newcommand{\sendrightauthtext}{\customarrownamedtext{-stealth}{ }{A}}
\newcommand{\sendrightsecure}{\customarrownamed{-stealth}{ }{S}}
\newcommand{\sendrightsecuretext}{\customarrownamedtext{-stealth}{ }{S}}

\newcommand{\sendleftrightauth}{\customarrownamed{stealth-stealth}{A}{A}}

\newcommand{\shortdots}{\kern-0.13em...\kern-0.03em}

\usepackage{bm}
\usepackage[boxed,vlined,dotocloa]{algorithm2e}
\crefname{algocf}{Algorithm}{Algorithms}
\DontPrintSemicolon
\SetAlgoInsideSkip{1em} 
\IncMargin{-0.8em} 
\SetAlCapSkip{0.5em} 
\SetCommentSty{myalgorithmecommfont}
\newcommand{\algo}[2]{{\textbf{Alg.:} \hypertarget{#1}{$\mathsf{#1}$\vspace{0.5em}}(#2)\\}}
\newcommand{\algoref}[1]{\hyperlink{#1}{\mathsf{#1}}}

\newcommand{\palgoref}[1]{\hyperlink{p-#1}{\mathsf{#1}}}
\newcommand{\palgodef}[1]{\hypertarget{p-#1}{\mathsf{#1}}}

\newcommand{\algoskip}{\vspace{0.8em}}

\newcommand{\return}[1]{\algoskip\Return{#1}}

\usepackage{adjustbox}

    \newcommand{\equals}{\leftarrow}
    \newcommand{\chooserandom}{\xleftarrow{r}}
    
    \newcommand{\choosematch}{\xleftarrow{\match}}
    \newcommand{\match}{.}

    \newcommand{\listof}[1]{\textbf{#1}}

    \newcommand{\sizeof}[1]{{|#1|}}

    \newcommand{\range}[2]{[#1, #2]}

    \newcommand{\C}{\mathcal{C}}    
    \newcommand{\plainC}{\mathcal{P}}    

    \newcommand{\Ncc}{m}           

    \newcommand{\PtC}{\textit{PtC}}
    \newcommand{\CtE}{\textit{CtE}}
    \newcommand{\CtVV}{\textit{CtVV}}
    \newcommand{\CtVVcc}{\textit{Ctvv}}

    \newcommand{\hCA}{\textit{hCA}}

    \newcommand{\stateVContent}{\Id, \ballot}

    \newcommand{\voteP}{P} 
    \newcommand{\voteC}{C} 
    \newcommand{\voteE}{E} 

    \newcommand{\listofvotes}[1]{\bm{#1}}
    
    \newcommand{\votesC}{\listofvotes{\voteC}}      
    \newcommand{\votesCSy}{\votesC_{\textit{Sync}}}
    \newcommand{\votesCCa}{\votesC_{\textit{Cast}}}
    \newcommand{\votesCCo}{\votesC_{\textit{Conf}}}
    \newcommand{\votesCAg}{\votesC_{\textit{Agre}}}

    \newcommand{\Id}{\textit{Id}} 
    \newcommand{\VV}{\textit{VV}} 
    \newcommand{\CA}{\textit{CA}} 
    \newcommand{\CV}{\textit{CV}} 

    \newcommand{\ballot}{\textit{B}}
    \newcommand{\ballotdef}{(\CtVV, \CA, \CV, \PtC)}

    \newcommand{\ccVV}{\textit{vv}}

    \newcommand{\ccCA}{\textit{ca}}
    \newcommand{\ccCV}{\textit{cv}}
    \newcommand{\ccvP}{\textit{p}}

    \newcommand{\ccballot}{\textit{b}}
    \newcommand{\ccballotdef}{(\CtVVcc, \ccCA, \ccCV, \ccvP)}

        \newcommand{\forallcc}{\forall \allcc}
        \newcommand{\allcc}{i \in \range{1}{\Ncc}}
        \newcommand{\forallothercc}{\forall \allothercc}
        \newcommand{\allothercc}{j \in \range{1}{\Ncc} \backslash i}

        \newcommand{\ccAND}{\oplus_{i=1}^\Ncc \hspace{0.2em}}
        
        \newcommand{\ccPROD}{\prod_{i=1}^\Ncc}

        \newcommand{\ccival}[2][i]{#2^{(#1)}}
        \newcommand{\ccjval}[2][j]{#2^{(#1)}}
        \newcommand{\cciballot}[1][i]{\ccival[#1]{\ccballot}}
        \newcommand{\cciballotdef}[1][i]{(\ccival[#1]{\CtVVcc}, \ccival[#1]{\ccCA}, \ccival[#1]{\ccCV}, \ccival[#1]{\ccvP})}

        \newcommand{\forallIds}{\forall \Id \in \listof{\Id}}

    \newcommand{\Za}{\mathbb{Z}_{n_a}} 
    \newcommand{\sizeofZa}{n_a} 
    \newcommand{\Zv}{\mathbb{Z}_{n_v}} 
    \newcommand{\sizeofZv}{n_v} 
    \newcommand{\Perm}{\mathbb{P}} 
    \newcommand{\PermC}{\Perm_\sizeof{\C}} 

    \newcommand{\hash}{\mathsf{H}}
    \newcommand{\signature}{\mathsf{S}}
    \newcommand{\encryption}{\mathsf{E}}

\newcommand*\circled[1]{\textcircled{\raisebox{-0.8pt}{#1}}}

\begin{document}
\title{Short Voting Codes For Practical Code Voting}
\author{Florian Moser\orcidID{0000-0003-2268-2367}}
\authorrunning{F. Moser}
\institute{INRIA Nancy, France
	\email{florian.moser@inria.fr}}

\maketitle
\begin{abstract}
	To preserve voter secrecy on untrusted voter devices we propose to use short voting codes. This ensures voting codes remain practical even if the voter is able to select multiple voting choices. 
	We embed the mechanism in a protocol that avoids complex cryptography in both the setup and the voting phase and relies only on standard cryptographic primitives.
	Trusting the setup, and one out of multiple server components, the protocol provides vote secrecy, cast-as-intended, recorded-as-cast, tallied-as-recorded, eligibility and universal verifiability. 
	
\end{abstract}

\section{Introduction}
In an internet voting system, the voter's client is usually considered untrusted. Consequentially, cast-as-intended and recorded-as-cast mechanism typically ensure the device does not alter the voter's choice. However, many of these mechanisms require the voter to enter their plain vote. Consequentially, the voter's device learns the plain vote of the voter.

A solution to safe-guard vote secrecy, even if the voter's client is malicious, are voting codes \cite{surevote2001}. The voter no longer enters their plain vote, but a voting code, with the voter's device unable to attribute the voting code to the plain vote it represents. However, voting codes often tend to be long, as they incorporate ciphertext of the plain vote or authentication secrets. This makes entering the codes tedious for the voter, even more so if multiple voting choices are chosen.

In this work, we introduce \textit{short} voting codes (1 or 2 digits in reasonable scenarios). Only as many different voting codes are needed as plain voting choices are available. In elections with a reasonable set of plain voting choices, the voting codes are therefore also of reasonable length. For each voter, the voting codes are assigned differently to the plain voting choices, so the voter's device does not learn the plain vote.

We implement this mechanism in a protocol providing vote secrecy, cast-as-intended, recorded-as-cast, tallied-as-recorded, eligibility verifiability and universal verifiability. We grant the adversary control of the network, the voter's device and some of the multiple server components we call control components. To guard our security properties, we assume the setup and one out of multiple server components trusted. Notably, our adversary and trust model, as well as the achieved properties, conform to the requirements set forward by the Swiss chancellery for Swiss political national elections \cite{oev2013}.\footnote{We note, and we have double-checked this fact with the Swiss Chancellery, that the current Swiss law and its derived ordinances do not forbid code voting.} Against an adversary only observing the exchange between the voter and the control components, the system achieves everlasting privacy.

The protocol relies on well-know cryptographic constructs, and tasks which users are familiar with from other applications (entering and comparing short codes, scanning QR-codes). The setup and voting phases need only few operations not more complex than signatures and hashes, and are efficient. The tally phase additionally needs a privacy-preserving tally mechanism such as a homomorphic tally or a verifiable shuffle, both well-understood mechanisms.

\paragraph{Related Work}
Voting codes to achieve privacy on an untrusted voters device's were first proposed in 2001 \cite{surevote2001}. Previous work directly uses the ciphertext as the voting code (as in BeleniosVS \cite{cortier2019beleniosvs}) or an identifier of the corresponding ciphertext (as in Pretty Understandable Democracy (PUD) \cite{prettyunderstandabledemocracy2013}). Further, some schemes use the voting codes to deliver additional guarantees. In Pretty Good Democracy, the voting codes are hard to guess to achieve receipt-freeness \cite{ryan2009pretty}. Many schemes also use the voting codes for authentication (i.e. a proposed code voting extension to the Swiss Post System \cite{volkamer2022increasing}, an other code voting scheme proposed in the Swiss setting \cite{hanni2023privintvot} and others \cite{surevote2001,helbach2007secure,joaquim2009veryvote,joaquim2013eviv,hanni2023privintvot}).

In our adversary and trust model, we observe that proposed protocols usually are complex, and involve purpose-specific cryptography. To extract the verification codes in the voting phase, the Swiss Post protocol requires two round trips between the control components involving multiple zero-knowledge proofs, encryptions and hashes \cite{postspecification2022}. For the same purpose, CHVote uses instead a novel oblivious transfer scheme \cite{haenni2017chvote}. A more recent proposal by Hänni et al. managed to simplify the voting phase using voting codes, however opted for a verifiable shuffle in the setup phase over all possible voting choices, with its associated complexity \cite{hanni2023privintvot}. In this work, we avoid the verifiable shuffle at the cost of an authenticated (but public, and therefore auditable) channel out of the trusted setup component.\footnote{Consider the extensions proposed in \cref{protocol:extensions:audit-setup}.}

\paragraph{Contributions}
We present a internet voting protocol using short voting codes. Only as many different voting codes are needed as voting choices are available.

We achieve vote secrecy without trust in the voter's device. Further, we provide cast-as-intended, registered-as-cast, tallied-as-recorded, eligibility verifiability and universal verifiability. Trust is required in the setup component, and a collection of control components, of which only one needs to be honest. The adversary may control the network and the untrusted components.

The protocol uses few, efficient and readily available cryptographic building blocks. Besides the verifiable tally mechanism, we need no more advanced cryptography than signatures, notably we do not need zero-knowledge proofs or homomorphic encryption in the setup and voting phases.

\section{Voter's view}
\begin{figure}
	\noindent\includegraphics[width=\textwidth]{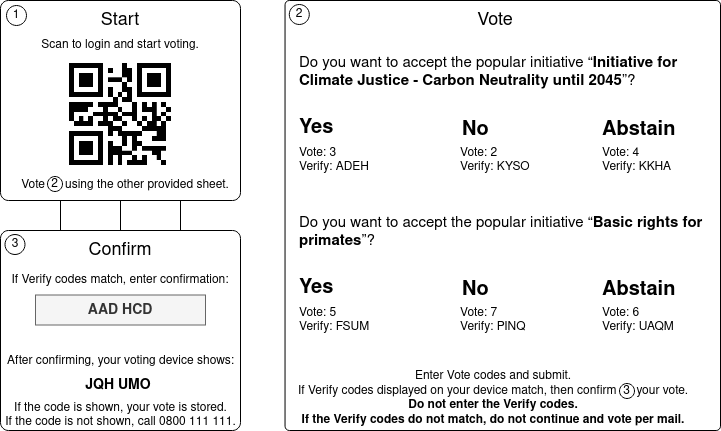}
	\caption{An example of a voting sheet using realistic parameters. \circled{1} contains a QR code with $\Id$. \circled{2} contains the voting codes $\voteC$ and their respective vote verifications $\VV_\voteC$. \circled{3} contains the $\CA$ (upper string) and the $\CV$ (lower string).}
	\label{formal-proto:voting_sheet}
\end{figure}

To introduce the protocol we present the voter's view. The voter is given a ballot sheet with a QR code to login, and codes to cast and confirm their vote. See \Cref{formal-proto:voting_sheet} for an example using realistic parameters of how this might look like. First, the voter scans the QR code to login \circled{1}. Then, the voter enters the short voting codes of the voting choices they intend to vote for \circled{2}. The voting codes are sent to the voting system, which responds with a verification code each. If and only if these codes match to what is printed on the voting sheet \circled{2}, the voter is instructed to confirm the vote \circled{3}. 

To perform their tasks as instructed, voters need to be able to scan QR codes and enter and compare short codes; all mechanisms already in use by existing large-scale systems (e.g. certificate scanning and second factor applications). To our knowledge, the usability of entering short voting codes in the voting context has not yet been explored. An extension of SPS incorporating (long) voting codes has however been shown to not reduce general usability \cite{volkamer2022increasing}. 

The voter application needs to essentially only forward values to the voting system and back to the user, and needs not to encrypt or sign the vote. The application may guide the user with basic validation (e.g. ensuring the voting code entered for the first question is between 2 and 4).

\section{Protocol}
\label{proposal:core-protocol}

The protocol is divided in three phases. In the setup phase, all parties receive required cryptographic material to run the voting and tally phase. In the voting phase, the voter submits their vote and performs their individual verifiability check. In the final tally phase, the votes are decrypted and counted, depending on the chosen tally mechanism.

\paragraph{Roles}
In the setup phase, the trusted \textit{Administrator} decides on the parameters of the election (participating voters, voting choices, length of keys, etc). The trusted \textit{Setup Component} then generates the corresponding key material and distributes it over secure channels to the other participants. 

In the voting phase, the \textit{Voter} casts and confirms their vote using their voting device over insecure channels. Voters can verify casting and confirmation is successful without trust in their device neither for privacy nor verifiability.

In all phases of the protocol, the \textit{Control Components} guarantee correctness and privacy of the election. We require only a single control component to be honest for the properties to be preserved. The \textit{Adversary} learns the data of untrusted protocol participants and can act on their behalf. Further, it can read, drop or add messages exchanged over untrustworthy channels.

\paragraph{Notation}
We choose random with $\chooserandom$. We denote modular addition with $\oplus_n$ for $n$ the modulo value. If $n$ is obvious from the context, we may omit it.

We use boldface for lists, for example $\listof{l} = [a_1, a_2, \shortdots]$. We call a list composed out of pairs a dictionary, for example $\listof{d} = [(a_1, b_1), (a_2, b_2)]$. We look for a match in a dictionary with $\choosematch$ (i.e. $x \choosematch (2, \match) \in [(1, a), (2, b)]$ results in $x = b$). If no match is found, or more than one, the process terminates.

We denote the group of permutations of integers up to $s$ as $\Perm_s$ (e.g. $\Perm_3$ includes $[1,2,3]$ or $[2,3,1]$). Using $*$, we apply permutations to themselves (e.g. $[1,2,3] * [2,3,1] = [3,1,2]$), and to the right value of the pairs in a dictionary (e.g. $[(1,a), (2,b), (3,c)] * [3,1,2] = [(1,b), (2,c), (3,a)]$).

When a party encounters an \textit{assert} with a falsy expression (like \textit{assert false} or \textit{assert 0}), then the party aborts processing.
We use $\sendrightauthtext$ and $\sendrightsecuretext$ to denote communication sent over the authenticated and secure channel, respectively.

\paragraph{Parameters}
The authorities decide on the following parameters and algorithms:

\begin{itemize}
	\item $\Za$ to pick authentication secrets from. The adversary is given their hashes, so $\sizeofZa$ must be big enough for picked values to be hard to brute-force.
	\item $\Zv$ to pick verification codes from. The adversary has a single try to convince the voter, so $\sizeofZv$ must be big enough for picked values to be hard to guess.
	\item $\listof{\Id}$ is a list of identifiers such that there is one for each eligible voter.\footnote{If an adversary guesses such an identifier, it may vote on the voter's behalf. While the voter will detect if a wrong vote has been cast, and consequentially does not confirm (which prevents tallying), they cannot use the system to cast their own vote anymore. Therefore, for availability, hard to guess identifiers are a good idea.}
	\item $\PtC$ (\textit{Plain to Codes}) is a dictionary which maps each plain vote $\voteP \in \plainC$ to a unique code $\voteC \in \C$ (e.g. $[(\textit{'Yes'}, A), (\textit{'No'}, B), (\textit{'Abstain'}, C)]$).
	\item $\hash$ is a pre-image resistant hash function (e.g. SHA256). The scheme needs to provide a hash function $\palgodef{Hash}(m)$ for message $m$.
	\item $\signature$ is an EUF-CMA secure signature scheme (e.g. ECDSA). The scheme needs to provide a sign function $\sigma \equals \palgoref{Sign}(sk, r, m)$ for secret key $sk$, randomness $r$ and message $m$. Further, the scheme needs to provide a verification function $\palgoref{Verify}(pk, m, \sigma) \in \{0,1\}$ for public key $pk$, message $m$ and signature $\sigma$. For a valid key pair $(pk, sk)$ it holds that $\forall r,m.  \palgoref{Verify}(pk, m, \palgoref{Sign}(sk, r, m)) = 1$. We write $\palgodef{Sign}(m)$ and $\palgodef{Verify}(m, \sigma)$ when the other arguments are clear from the context.
	\item $\encryption$ is an IND-CPA secure public key encryption scheme (e.g. ElGamal) to produce ciphertext suitable as input for the privacy-preserving tally. The scheme needs to provide an encryption function $E \equals \palgoref{Enc}(pk, r, m)$ for $pk$ the public key, $r$ the encryption randomness and $m$ the message. Likely, $\palgoref{Enc}$ operates on an aggregated public key of which a private key share is at each control component. We write $\palgodef{Enc}(m)$ when the other arguments are clear from the context.
\end{itemize}

\subsection{Setup}
\label{proposal:core-protocol:setup}

In the setup phase, the setup component generates key material and sends the result to the control components and to the voters.

\noindent
\begin{minipage}{.38\linewidth}
	\begin{algorithm}[H]
		\algo{GenPartialBallot}{}
		
		\For{$\voteC \in \C$}{
			$\ccVV \chooserandom \Zv$\;
			$\CtVVcc \equals \CtVVcc \cup (\voteC, \ccVV)$\;
		}
		
		\algoskip
		$\ccCA \chooserandom \Za$\;
		$\ccCV \chooserandom \Zv$\;
		
		\algoskip
		$\ccvP \chooserandom \PermC$\;
		
		\return{$\ccballotdef$}\;
		\caption[$\mathsf{GenPartialBallot}$]{Generates a partial ballot.}
		\label{alg:GenPartialBallot}
	\end{algorithm}
\end{minipage}
\begin{minipage}{.62\linewidth}
	\begin{algorithm}[H]
		\algo{MergePartialBallots}{$\forallcc.\ccival{\ccballot}$}
		$\forallcc . \cciballotdef \equals \ccival{\ccballot}$
		
		\algoskip
		\For{$\voteC \in \C$}{
			$\forallcc . \ccival{\ccVV} \choosematch (\voteC, \match) \in \ccival{\CtVVcc}$\;
			$\CtVV \equals \CtVV \cup (\voteC, \ccAND \ccival{\ccVV})$\;
		}
		
		\algoskip
		$\CA \equals \ccAND \ccival{\ccCA}$\;
		$\CV \equals \ccAND \ccival{\ccCV}$\;
		
		\algoskip
		$\PtC \equals  \PtC * \ccPROD \ccival{\ccvP}$\;
		
		\return{$\ballotdef$}\;
		\caption[$\mathsf{MergeBallot}$]{Merges partial ballots generated by \cref{alg:GenPartialBallot}.}
		\label{alg:MergePartialBallots}
	\end{algorithm}
\end{minipage}

\begin{protocol}{Setup phase where the setup component establishes key material for the control components (CC) and the voters.}{formal-proto:setup-phase--establish-key-material}
	\begin{tabular*}{\textwidth}{ l c l @{\extracolsep{\fill}} c r }
		\textbf{Voter}    & & \textbf{Setup component}    & & $\ccival{\textbf{CC}}$  \\
		$\forallIds$      & &    $\forallIds$             & & $\allcc$   \\ \hline \\[-0.5em]
		
		& & $\forallcc. \cciballot \chooserandom \algoref{GenPartialBallot}()$                      & & \\
		& & $\ballot \equals \algoref{MergePartialBallots}(\{\cciballot \mid \allcc\})$                & & \\
		
		& $\stateVContent$    & & $\Id, \ccival{\ccballot}$ \\[-0.8em]
		& \sendleftsecure     & & \sendrightsecure 		    \\[2em]
		
		& & $(\dots, \CA, \dots, \PtC) \equals \ballot$  \\
		& & $\hCA \equals \palgoref{Hash}(\CA)$  \\
		& & $\CtE \equals \{(\voteC, \voteE) \mid \voteC \in \C; \voteE \equals \palgoref{Enc}(\voteC * \PtC^{-1})\}$  \\
		
		& & & $\hCA, \CtE$    \\[-0.8em]
		& & & \sendrightauth           \\[-0.2em]
		
	\end{tabular*}
\end{protocol}

\subsection{Voting phase}
\label{proposal:core-protocol:voting}

In the voting phase, the voter casts and confirms their vote together with the control components (see \cref{formal-proto:voting-phase-cast-vote} (cast vote) and \cref{formal-proto:voting-phase-confirm-vote} (confirm vote)).

\begin{protocol}{Voting phase (1/2) where the voter casts their vote. The control components (CC) check the authentication, synchronize state and then respond with the verification specific to the received vote. Finally, the voter checks the verification.}{formal-proto:voting-phase-cast-vote}
	\begin{tabular*}{\textwidth}{ l @{\extracolsep{\fill}} c l c r }
		\textbf{Voter} & & $\ccival{\textbf{CC}}$ & & $\ccjval{\textbf{CC}}$  \\
		$\forallIds$ & &                 $\allcc$              & & $\allothercc$  \\ \hline \\[-0.5em]
		
		knows $\stateVContent$  & & knows $(\Id, \ccival{\ccballot}, \CtE)$ \\
		$(\CtVV, \shortdots, \PtC) \equals \ballot$ & & $(\CtVVcc, \shortdots) \equals \ccival{\ccballot}$ \\[1em]
		 
		decides plain vote $\voteP$ \\
		$\voteC \choosematch (\voteP, \match) \in \PtC$  \\
		
		& $\Id, \voteC$ \\[-0.8em]
		& \sendright    \\[-0.2em]
		
		& & asserts $\voteC \in \C$ \\
		& & asserts $\Id \notin \votesCSy$ \\[1em]
		
		& & $\votesCSy \equals \votesCSy \cup \{\Id\}$ \\
		& & $\voteE \equals (\voteC, \match) \in \CtE$ \\
		& & $\ccival{\sigma} \equals \palgoref{Sign}(\Id, \voteE)$ 	\\[0.5em]
		
		& & & \multicolumn{2}{c}{$\ccival{\sigma}$} \\[-1em]
		& & & \multicolumn{2}{c}{$\sendleftrightauth$} \\[-0.6em]
		& & & \multicolumn{2}{c}{$\ccjval{\sigma}$} \\[1.5em]
		& & \multicolumn{3}{l}{assert $\forallothercc.\palgoref{Verify}((\Id, \voteE), \ccjval{\sigma})$} \\[1em]
		
		& & $\votesCCa \equals \votesCCa \cup \{(\Id, \voteE)\}$ \\
		& & $\ccival{\ccVV} \choosematch (\voteC, \match) \in \CtVVcc$ \\
		
		& $\ccival{\ccVV}$ \\[-0.7em]
		& \sendleft \\
			
		$\VV \choosematch (\voteC, \match) \in \CtVV$ \\
		asserts $\VV = \ccAND \ccival{\ccVV}$                         & & \\
	\end{tabular*}
\end{protocol}

\begin{protocol}{Voting phase (2/2) where the voter confirms their vote. The control components (CC) check the authentication and then respond with a verification that the confirmation has been received. Finally, the voter checks the verification.}{formal-proto:voting-phase-confirm-vote}
	\begin{tabular*}{\textwidth}{ l @{\extracolsep{\fill}} c r }
		\textbf{Voter}       & & $\ccival{\textbf{CC}}$  \\
		$\forallIds$         & & $\allcc$                \\ \hline \\[-0.5em]
		
		knows $\stateVContent$  & & knows $(\Id, \ccival{\ccballot}, \hCA)$ \\
		$(\shortdots, \CA, \CV, \shortdots) \equals \ballot$  & & $(\shortdots, \ccival{\ccCV},\shortdots) \equals \ccival{\ccballot}$ \\
		
		& $\Id, \CA$    \\[-0.8em]
		& \sendright    \\
		
		& & asserts $(\Id, \match) \in \votesCCa$ \\
		& & asserts $\palgoref{Hash}(\CA) = \hCA$ \\[1em]
		
		& & $\voteE \equals (\Id, \match) \in \votesCCa$ \\
		& & $\votesCCo \equals \votesCCo \cup \{(\Id, \voteE)\}$ \\
		
		& $\ccival{\ccCV}$   \\[-0.7em]
		& \sendleft \\
		
		asserts $\CV = \ccAND \ccival{\ccCV}$                                  & & \\
	\end{tabular*}
\end{protocol}

If the voter reaches the end of both protocols, the voting procedure was successful and no other actions are necessary. If otherwise the voter aborts, they are instructed to use a different voting channel.

The system includes the vote for tallying as soon as the voter entered the confirm authentication $\CA$. But the voter may also participate over a different voting channel (e.g. because they did not receive a correct confirm verification), so duplicates over the different voting channels need to be dealt with. 

\subsection{Tally phase}
\label{proposal:core-protocol:tally}

At the end of the voting phase, each control component has a list of confirmed votes $\votesCCo$ which need to be tallied. All correctly processing control components have the same $\votesCCo$ in their local state. However, dishonest control components might have added, modified or dropped entries in their local $\votesCCo$, so we need to clearly define what list of ciphertext is to be tallied.

\begin{definition}
\label{definition:votescag}
$\votesCAg$ contains all ciphertext entries $\voteE$ for which it holds that for some voter $\Id$, the $\CA$ is known and from all control components there exists a signature over $(\Id, \voteE)$ (i.e. $\forallcc . \exists \ccival{\sigma} . \palgoref{Verify}((\Id, \voteE), \ccival{\sigma}$).
\end{definition}
 
To establish $\votesCAg$, each control component sends $\votesCCo$ with the proof for each entry ($\CA$ and signatures over $(\Id, \voteE)$) to all other control components. Each control component then adds all $\voteE$ to $\votesCAg$ for which the definition holds, considering their own $\votesCCo$ and proofs, as well as the $\votesCCo'$ and proofs received from the other control components.

After establishing $\votesCAg$, the control components then execute the privacy-preserving verifiable tally-mechanism, with the ciphertext given by $\votesCAg$ as input. The correct execution of all the steps are provable to third parties.

\subsection{Extensions}
\label{proposal:extensions}

With the core protocol given, we now improve on functionality and security.

\paragraph{Practical availability}
As presented here, the voter directly communicates with the control components, hence an adversary may try to flood the control components with bogus requests. As soon as a single control component is unable to process more requests, honest voters can no longer cast and confirm votes. However, verification of authentication and validation of votes is efficient: Only hash and set membership checks are needed (but e.g. no asymmetric cryptography). Further, the checks operate solely on public data. 

To improve availability, an untrusted server component can be introduced specifically hardened to resist such flooding attacks. The server component is placed in between the voter and the control components. If said system then operates correctly, any request which reaches the control components will pass authentication and validation, minimizing load on the control components' side.

\paragraph{Supporting multiple elections, voting choices and eligibility}
So far, we only described how a single popular vote or election will proceed for a single voting choice. However, votes are usually held over multiple issues at the same time. Further, depending on the issue voted on, multiple voting choices are possible. Also, voters might not all be of the same eligibility. To ensure the protocol is applicable to many voting scenarios, we aim to support k-out-of-n elections with varying eligibility per voter, which for example corresponds to elections in Switzerland \cite[Chapter 2.2.2/2.2.3]{haenni2017chvote}.

To support multiple issues at the same time, additional sets of voting codes $\C$ are chosen. Consequentially, the setup phase handles additional permutations corresponding to each additional set of voting codes.
To support k-out-of-n issues, the voter has to submit exactly $k$ voting options out of $\sizeof{\C} = n$. 
To support different eligibilities, voters are restricted in the sets of voting codes they can submit a vote for. 
Consequentially, the vote validation (see \cref{formal-proto:voting-phase-cast-vote}) is extended to ensure the sets of voting codes submitted, and the number of submitted voting options per set of voting codes, is valid.

To improve privacy of voters with restricted eligibility, the tally mechanism may tally each issue (i.e. each set of voting options) separately. 

\paragraph{Audit of the setup component}
\label{protocol:extensions:audit-setup}
The setup component needs to be fully trusted. Notably it could switch vote and verification codes of plain votes, such that a different voting option is voted for than intended. We can introduce an audit procedure to improve detection of a misbehaving setup component.

To make an audit of the setup component practical, we first ensure it only runs deterministic algorithms. We implement this by running $\algoref{GenPartialBallot}()$ at each control component instead of the setup component, and consequentially reverse the corresponding secure channel (now from the control components to the setup component). Note how $\algoref{MergePartialBallots}()$ ensures that if at least one $\ccival{\ccballot}$ is generated honestly, the adversary has no advantage to guess values in $\ballot$.

To audit the setup component, we add $n$ additional entries to $\listof{\Id}$, for $n$ chosen in such a way that if $n$ voters are audited, the risk is sufficiently minimized. After the setup component finishes execution, each control component $i$ chooses some subset of voters to audit $\ccival{\listof{\Id}_A} \subset \listof{\Id}$ for $\sizeof{\cup \ccival{\listof{\Id}_A}} = n$. Any corresponding $\Id$ can no longer be used to cast votes, but instead the control components publish $\ccival{\ccballot}$. The control components then assert that $\ballot$, $\hCA$ and ${(\voteC, \voteE)}$ have been generated correctly.

\paragraph{Making the protocol robust outside the adversary model}
As presented, the protocol is secure within the adversary model. However, we may also want to preserve security properties outside the adversary model. So even if a presumed honest party is controlled by the adversary, given some additional assumptions, we still want to deliver some guarantees. 

Note how the mechanisms presented here cannot prevent all attacks of a presumed honest party. For example, no mechanism prevents a dishonest setup component to person-in-the-middle the individual verifiability check, and if all control components are dishonest, they can easily drop votes. But the mechanisms might still manage to increase security in practical scenarios.

To prevent a dishonest setup component to insert votes for abstaining voters, the control components can inform the user directly about whether their vote is considered in the tally (i.e. is part of $\votesCCo$) after the voting phase has ended. This assumes an authenticated channel to the voter for that purpose.

To prevent a dishonest setup component to target specific voters (i.e. switching voting and verification codes for voters likely voting in a certain way), the secure channel to the voter can obfuscate to the setup component which actual voter (name, address) is assigned which $\Id$. For example, if the secure channel to the voter is implemented over postal mail, the voting sheets can be printed first, physically shuffled, and only then be put into addressed envelopes.

To prevent jointly dishonest control components to insert or modify votes, the setup component can generate a signature key pair per voter, of which the public key is published. The voter device scans the private key together with the identification, and signs the submitted vote.\footnote{Note how this does not change the voter interaction.} Only votes which have a valid signature are included in the tally, to be checked by an auditor. This assumes an honest setup component, an honest voter device and an honest auditor.

To prevent jointly dishonest control components to learn the vote of a voter, an additional tally component can be introduced which contributes part of the decryption key and participates in the tally. The tally component is therefore included in the group of control components of which only one needs to be honest for the security properties of the tally mechanism to be preserved. From a practical point of view, the tally component may indeed be easier to secure, as it does not need to participate in the voting phase (hence needs not be "online"). If this same tally component is involved in proving the participation of voters, it further ensures that dishonest control components cannot drop or add votes.

\bibliographystyle{splncs04}
\bibliography{references}

\end{document}